\documentclass[a4paper,11pt]{article}
\usepackage{pos}

\title{Towards a Determination of thermal static Potential at Finite Density in (2+1)-flavor QCD}
\ShortTitle{Thermal static Potential at Finite Density in (2+1)-flavor QCD}
\newcommand \hmu {\hat{\mu}}

\author*[a]{Jishnu Goswami}
\author[a]{Dibyendu Bala}
\author[a]{Olaf Kaczmarek}

\affiliation[a]{Fakult\"at f\"ur Physik, Universit\"at Bielefeld, D-33615 Bielefeld,
Germany}

\emailAdd{jishnu@physik.uni-bielefeld.de}
\emailAdd{dibyendu.bala@physik.uni-bielefeld.de}
\emailAdd{okacz@physik.uni-bielefeld.de}

\abstract{We study the thermal static potential for (2+1)-flavor QCD at nonzero density through a Taylor expansion around vanishing chemical potentials. From Taylor expanded Wilson line correlators, we extract the $\hmu^2$ coefficient of the real and imaginary part of the potential in light and strange flavor channels and in the baryon number and electric charge channels. We observe an enhancement of in-medium screening at intermediate and large separations. The effect is visible in both the real and imaginary parts to the extracted $\hmu^2$ contribution of the static potentials and  provides a first step toward constraining in-medium heavy-quark interactions relevant for the Beam Energy Scan program at RHIC and future FAIR experiments.}
\FullConference{The 42nd International Symposium on Lattice Field Theory (LATTICE2025)\\
2-8 November 2025\\
Tata Institute of Fundamental Research, Mumbai, India\\}

\begin{document}
\maketitle
\section{Introduction}
Understanding the fate of the heavy quark bound states in hot and dense strongly interacting matter, formally known as quark gluon plasma (QGP) created in heavy ion collisions remains one of the central goals in high energy nuclear physics. Heavy quarkonia, such as charmonium and bottomonium, are valuable probes of the quark-gluon plasma (QGP) produced in heavy-ion collisions at RHIC and the LHC. This idea goes back to the seminal suggestion of Matsui and Satz \cite{Matsui:1986dk}, where color screening in the deconfined medium leads to the suppression of quarkonium yields in heavy-ion collisions. The quark mass $M_q$ is much larger than the medium temperature $T$ and the strong interaction scale $\Lambda_{\text{QCD}}$. Therefore, the dynamical behavior of quarkonia in the medium can be described by a Schrödinger equation with a complex thermal potential, where the real part encodes color screening, while the imaginary part accounts for the collisional effects of medium partons on quarkonia
\cite{Laine:2006ns, Burnier:2007qm, Brambilla:2008cx}. This complex thermal potential is defined in the static limit through the analytic continuation of the Wilson loop in Euclidean time($\tau$), $\lim_{t\to \infty}W(r,\tau\to it)\sim e^{-i V_{T}(r)t}$. The extraction of the thermal potential from noisy Euclidean lattice data is an ill-posed problem, and thus further physics input is required for a reliable extraction through analytic continuation. On the lattice, this has been studied extensively using Wilson-line correlators fixed in Coulomb gauge and using Bayesian techniques or physically motivated constraints on the analytic structure of the correlation function \cite{Burnier:2014ssa, Burnier:2015tda,Bala:2019cqu,Ali:2025iux,Bala:2026vnl}. These studies have established that both the real and imaginary parts of the potential are modified in the thermal medium. Extending this analysis to nonzero density is of particular interest for understanding quarkonium behavior in low beam energy experiments at RHIC and in future experiments at FAIR. The QCD equation of state is reliably obtained with a controlled Taylor expansion for small values of the baryon chemical potential, $\mu_{B}/T \lesssim 2$ in \cite{Gavai:2008zr, Bazavov:2017dus,Bollweg:2022rps,Bollweg:2022fqq}. Following this, we also perform a Taylor expansion of the Wilson-line correlator at finite chemical potential up to $\hat\mu^2$, and hence extract corrections to the thermal static potential up to second order.
We analyze both quark-flavor and conserved-charge chemical-potential channels and determine the corresponding corrections to the real and imaginary parts of the potential near the pseudo-critical temperature.

The main goal of this work is to quantify how finite density modifies the in-medium heavy-quark interaction in $(2+1)$-flavor QCD. Our analysis shows that the $\hat\mu^2$ contribution enhances medium effects at intermediate and large separations, providing the first lattice constraints on the finite-density dependence of the thermal static potential.

\section{Lattice action, simulation details, and Wilson-line correlators at nonzero chemical potentials}
We use the $(2+1)$-flavor highly improved staggered quark (HISQ) action, with the partition function
\begin{align}
Z &= \int DU \, e^{-S_g}\prod_{f=u,d,s}\det M_f(m_f,\hmu_f), \nonumber\\
{}&M_f(m_f,\hmu_f) = D_{\rm HISQ}(\hmu_f)+m_f .
\label{eq:partitionfunc}
\end{align}
Throughout this work we use degenerate light quark masses, $m_u=m_d$ and $\hmu_f=\mu_f/T$. The calculations are performed on $64^3\times 16$ lattices with a mass ratio on the line of constant physics, $m_l/m_s=1/27$, where $m_s$ (the strange quark mass) tuned to its physical values, corresponding to physical pion mass. The temperature is fixed to $T=151.92~\mathrm{MeV}$, which is close to the pseudo-critical temperature, $T_{pc}=156.5(1.5)~\mathrm{MeV}$~\cite{HotQCD:2018pds}, of $(2+1)$-flavor QCD. We have about $7014$ gauge configurations separated by $10$ trajectories.
\begin{figure}[t]
    \centering
    \includegraphics[width=0.44\textwidth]{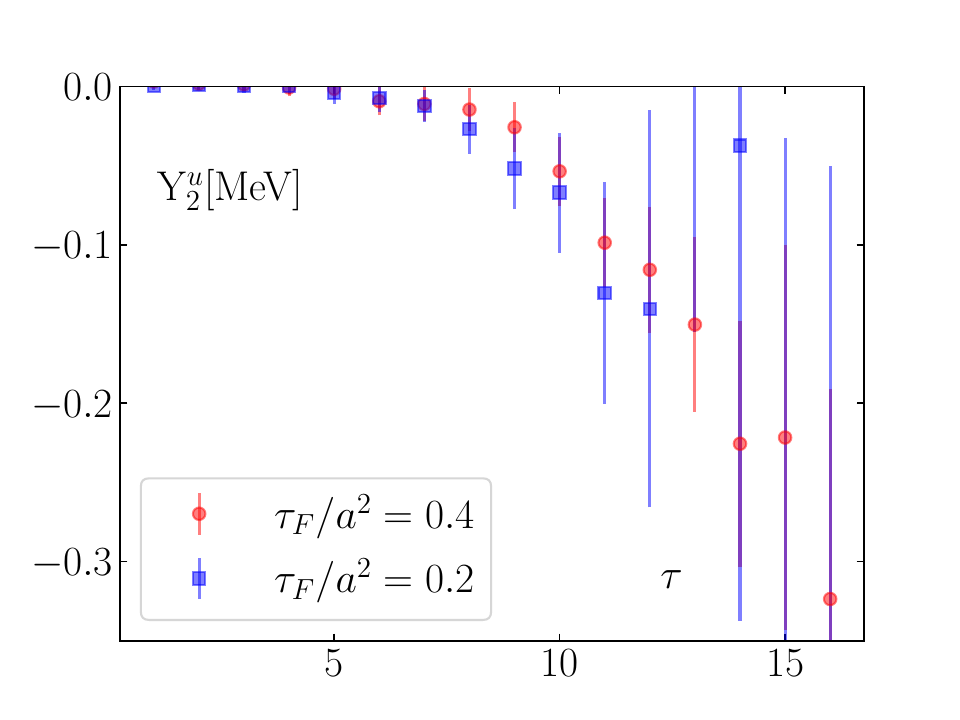}
    \includegraphics[width=0.44\textwidth]{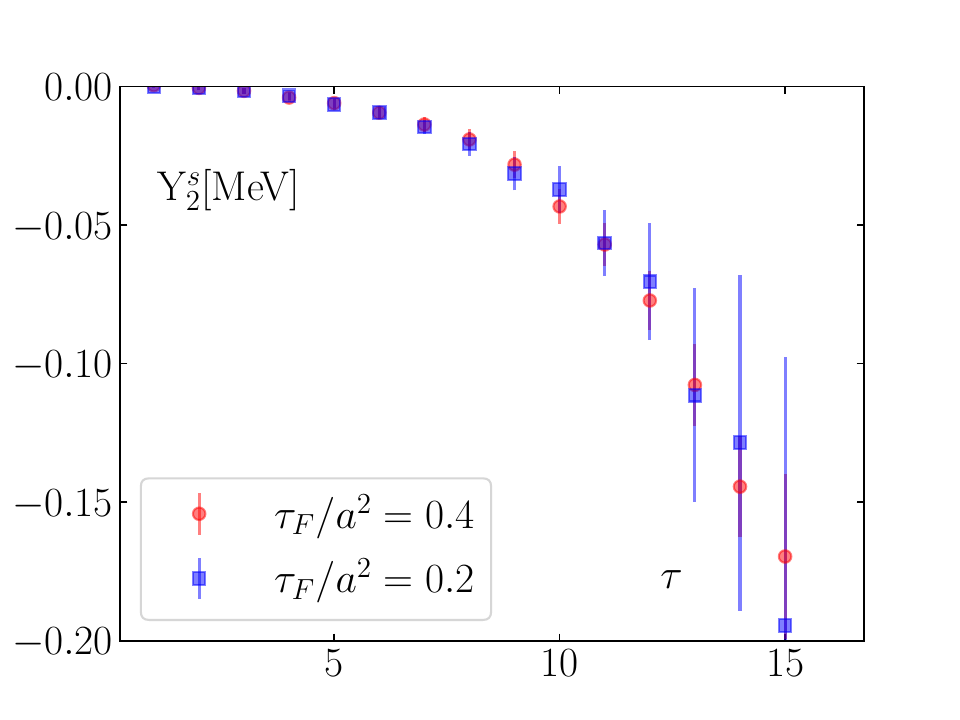}
    \caption{Second order Taylor expansion coefficient $Y^u_2$ (left) and $Y^s_2$ (right) of the Wilson line correlator, shown as a function of Euclidean time separation $\tau$ at fixed spatial separation $r=1~\rm{fm}$ and temperature $T = 151.92~\rm{MeV}$ for two flow times,$\tau_f/a^2=0.2$ and $0.4$.}
    \label{fig:W2_W0}
\end{figure}
\subsection{Wilson-line correlator at nonzero chemical potentials}
Let $W(r,\tau;\hat\mu_X)$ denote the Wilson-line correlator in Coulomb gauge at spatial separation $r$ and Euclidean time separation $\tau$ in the presence of a chemical potential $\hat\mu_X$. Here, $X$ may denote either a quark-flavor channel, $u,d,s$, or a conserved-charge channel, $B,Q,S$. Expanding around vanishing chemical potentials, the correlator can be written in the flavor basis as
\begin{align}
W(r,\tau;\hat\mu_u,\hat\mu_d,\hat\mu_s)
&= W_0(r,\tau)
+ \sum_{i+j+k=\,2}
\frac{1}{i!\,j!\,k!}\,
W_{ijk}^{uds}(r,\tau)\,
\hat\mu_u^i \hat\mu_d^j \hat\mu_s^k
+ \mathcal{O}(\hat\mu^4),
\nonumber\\
W_{ijk}^{uds}(r,\tau)
&=
\frac{\partial^{\,i+j+k}W(r,\tau;\hat\mu_u,\hat\mu_d,\hat\mu_s)}
{\partial\hat\mu_u^i\,\partial\hat\mu_d^j\,\partial\hat\mu_s^k}
\Bigg|_{\hat\mu_u=\hat\mu_d=\hat\mu_s=0}.
\label{eq:W_taylor_uds}
\end{align}
The same expansion can equivalently be expressed in the conserved-charge basis,
\begin{equation}
W(r,\tau;\hat\mu_B,\hat\mu_Q,\hat\mu_S)
=
W_0(r,\tau)
+ \sum_{i+j+k=\,2}
\frac{1}{i!\,j!\,k!}\,
W_{ijk}^{BQS}(r,\tau)\,
\hat\mu_B^i \hat\mu_Q^j \hat\mu_S^k
+ \mathcal{O}(\hat\mu^4).
\label{eq:W_taylor_BQS}
\end{equation}
Odd-order terms vanish because of the symmetry under
$\hat\mu_X \to -\hat\mu_X$.
Since the linear term is absent, the logarithm of the correlator at $\mathcal{O}(\hat\mu^2)$ takes the simple form
\begin{equation}
\ln W(r,\tau;\hat\mu_u,\hat\mu_d,\hat\mu_s)
=
\ln W_0(r,\tau)
+ \sum_{i+j+k=2}
Y_{ijk}^{uds}(r,\tau)\,
\hat\mu_u^i \hat\mu_d^j \hat\mu_s^k
+ \mathcal{O}(\hat\mu^4),
\label{eq:logW_taylor}
\end{equation}
where we define the curvature coefficients
\begin{equation}
Y_{ijk}^{uds}(r,\tau)
=
\frac{1}{i!\,j!\,k!}\,
\frac{W_{ijk}^{uds}(r,\tau)}{W_0(r,\tau)},
\qquad i+j+k=2.
\label{eq:Y2_ratio}
\end{equation}
For brevity, we omit vanishing indices in the notation, e.g.,
$Y_{200}^{uds}\equiv Y_2^u$, $Y_{110}^{uds}\equiv Y_{11}^{ud}$, and
$Y_{101}^{uds}\equiv Y_{11}^{us}$. Likewise,
$W_{200}^{uds}\equiv W_2^u$, $W_{110}^{uds}\equiv W_{11}^{ud}$, and
$W_{101}^{uds}\equiv W_{11}^{us}$.

$W_0(r,\tau)$ is the ensemble average of the Wilson-line correlator at vanishing chemical potentials, while the second-order coefficients are obtained from the corresponding second-derivative estimators in the selected channel. These coefficients can be expressed in terms of the operators
\begin{equation}
D_n^f
\equiv
\left.
\frac{\partial^n}{\partial \hat\mu_f^{\,n}}
\ln \det M_f(m_f,\hat\mu_f)
\right|_{\hat\mu_f=0},
\qquad f\in\{u,d,s\},
\label{eq:Dn}
\end{equation}
which, as discussed in Ref.~\cite{Doring:2005ih,Allton:2005gk}, can be written as traces involving inverse Dirac matrices and their derivatives with respect to the chemical potentials.

We estimate these traces stochastically using Gaussian random noise vectors. Among the required contributions, $(D_1^u)^2$ is the noisiest term in our calculation. For this term, we use $2000$ random noise vectors, while for all remaining traces we use $500$ random noise vectors.
\begin{figure}[t]
\centering
\includegraphics[width=0.46\textwidth]{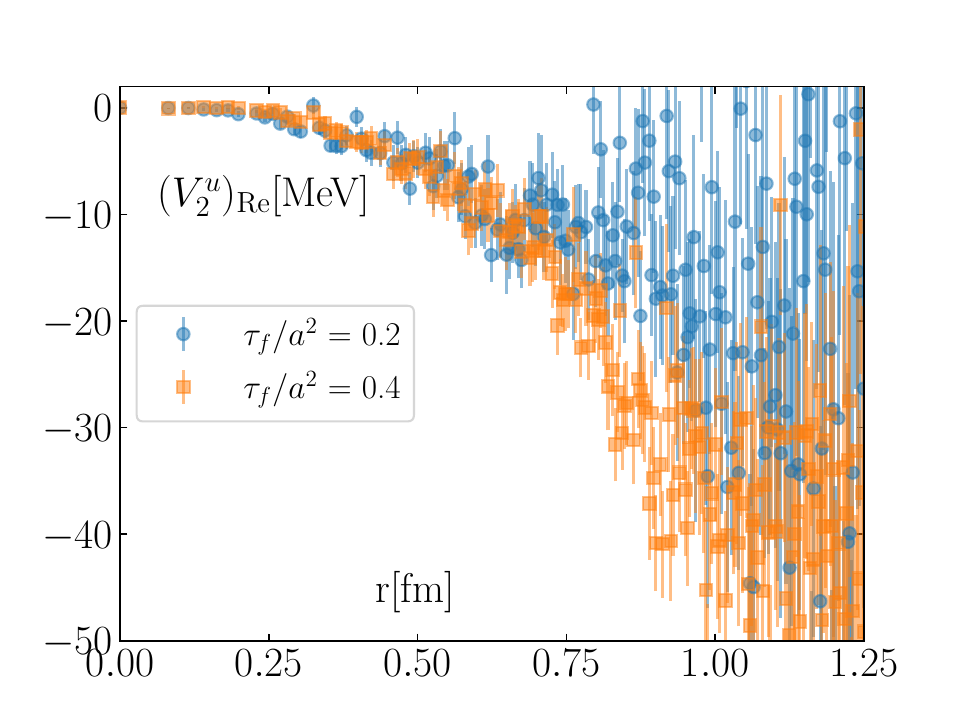}
\includegraphics[width=0.46\textwidth]{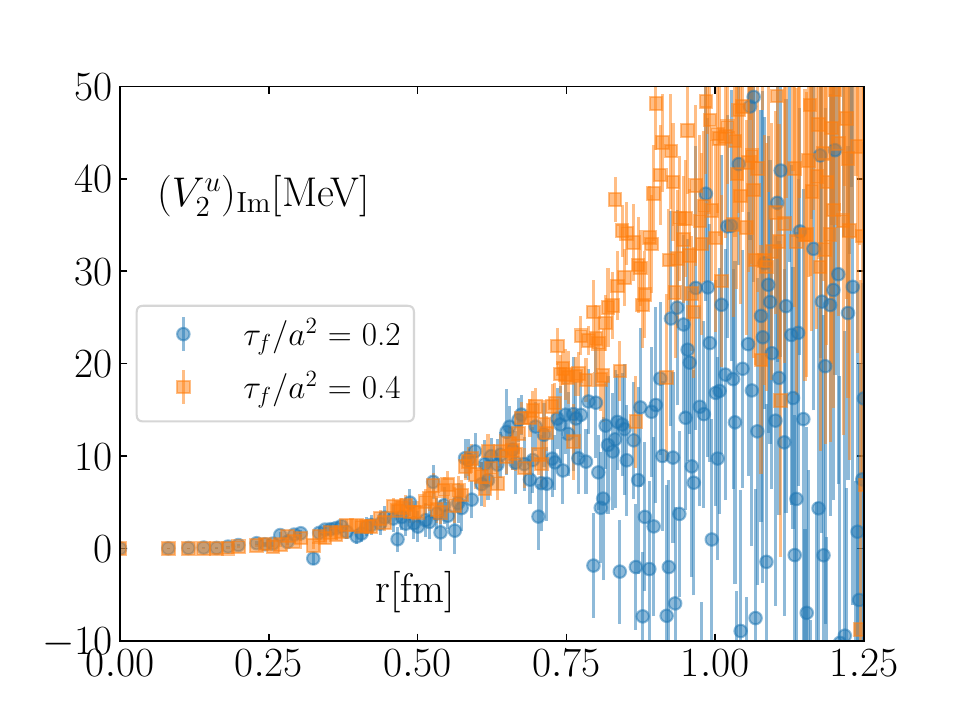}
\includegraphics[width=0.46\textwidth]{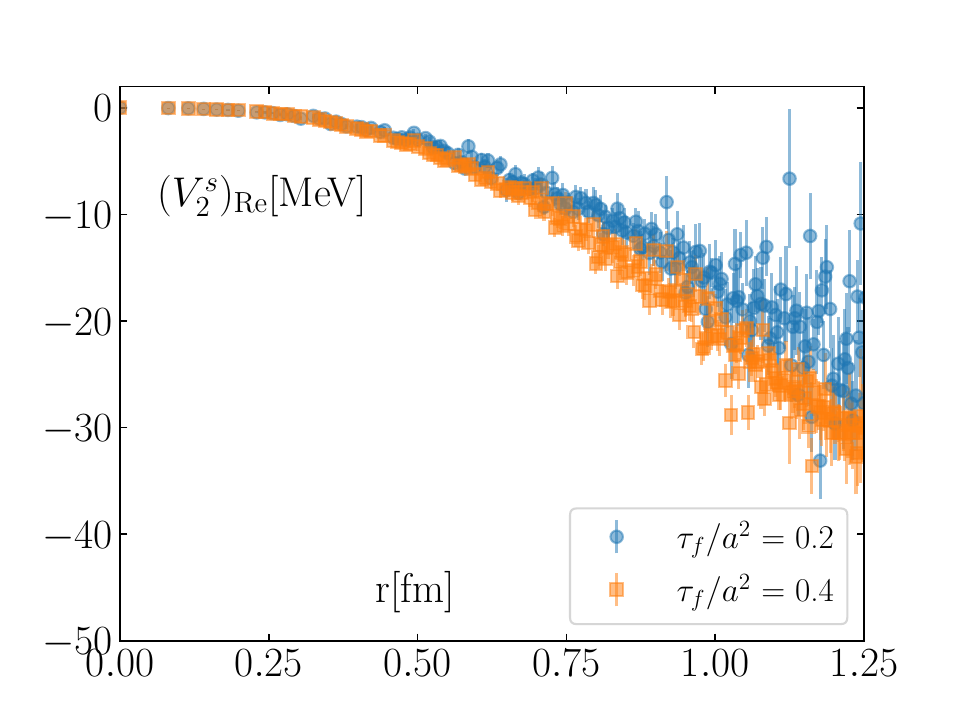}
\includegraphics[width=0.46\textwidth]{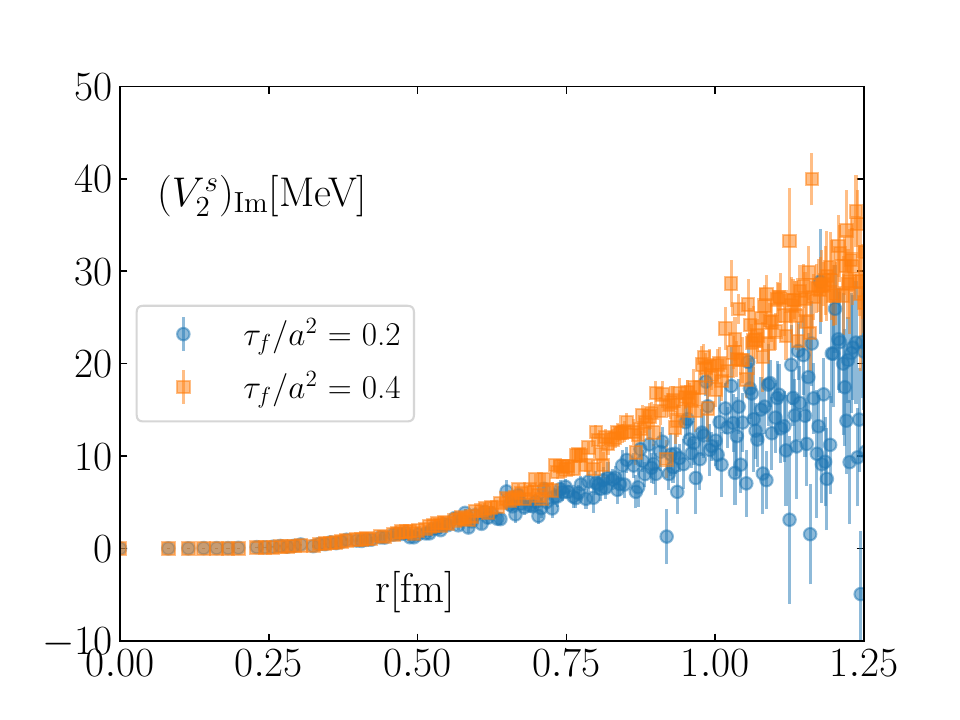}
\caption{Second-order Taylor-expansion coefficients of the thermal static potential in the flavor basis. Shown are $(V_2^{u})_{\mathrm{Re}}$, $(V_2^{u})_{\mathrm{Im}}$, $(V_2^{s})_{\mathrm{Re}}$, and $(V_2^{s})_{\mathrm{Im}}$ as functions of the separation $r$ at $T=151.92~\mathrm{MeV}$ for $\tau_f/a^2=0.2$ and $0.4$.}
\label{fig:V_mu_Re}
\end{figure}
\begin{figure}[htb]
\centering
\includegraphics[width=0.45\textwidth]{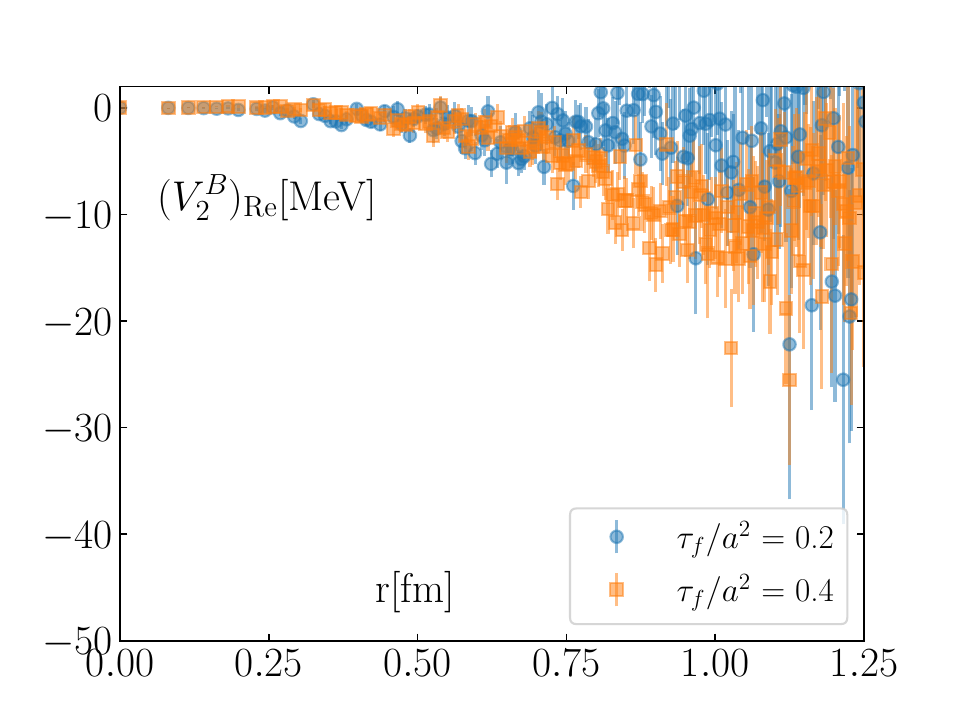}
\includegraphics[width=0.45\textwidth]{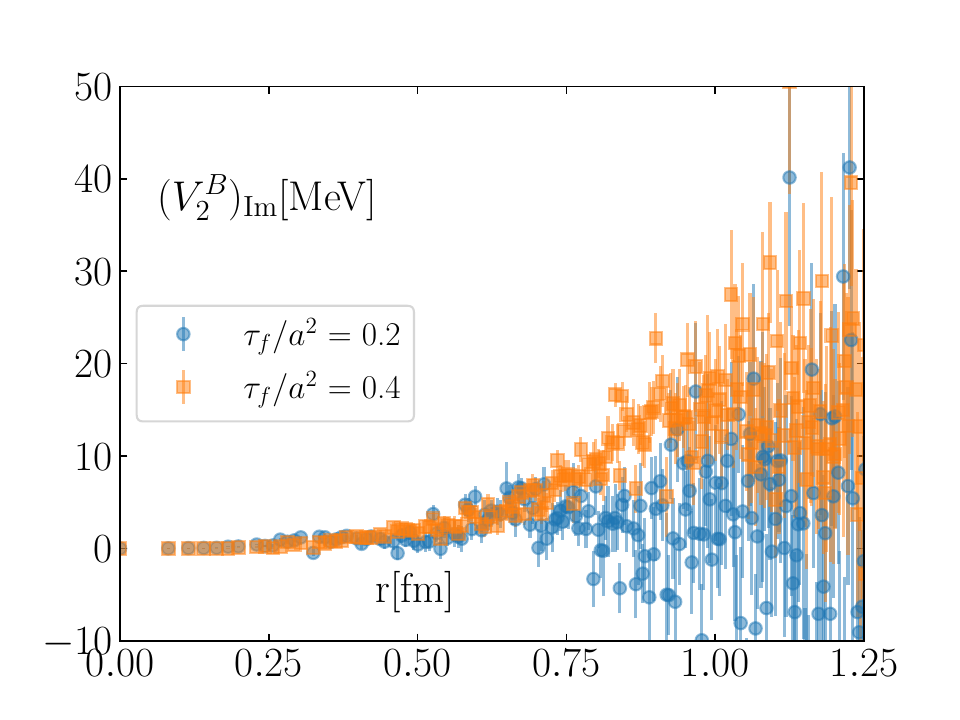}
\includegraphics[width=0.45\textwidth]{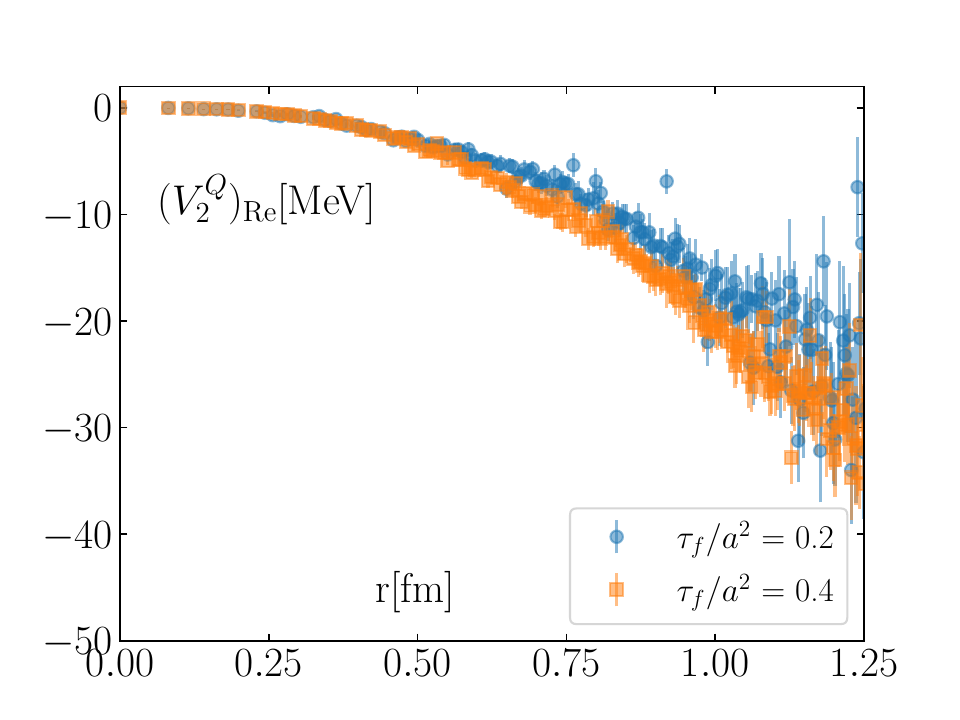}
\includegraphics[width=0.45\textwidth]{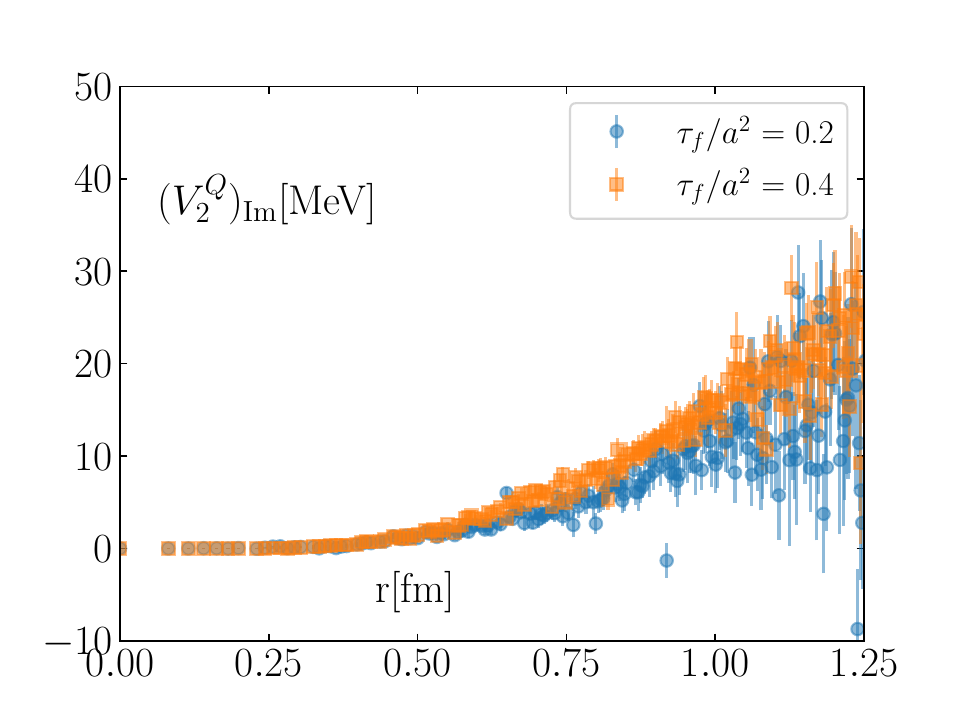}
\caption{Second-order Taylor-expansion coefficients of the thermal static potential in the conserved-charge basis. Shown are $(V_2^{B})_{\mathrm{Re}}$, $(V_2^{B})_{\mathrm{Im}}$, $(V_2^{Q})_{\mathrm{Re}}$, and $(V_2^{Q})_{\mathrm{Im}}$ as functions of the separation $r$ at $T=151.92~\mathrm{MeV}$ for $\tau_f/a^2=0.2$ and $0.4$.}
\label{fig:V_mu_im}
\end{figure}
\section{Potential fitting ansatz and $\hmu^2$ parameterization}
As mentioned above, the analytic continuation lattice Euclidean data has to be  guided through physics input, one such parametrization that has been widely used, ~\cite{Bala:2019cqu,Bala:2020tdt} is given by,
\begin{equation}
W(r,\tau)
=
A_0(r)\,
\exp\!\Big[
- V_0^{Re}(r)\,\tau
-\frac{N_\tau V_0^{Im}(r)}{\pi}\ln\sin\!\Big(\frac{\pi\tau}{N_\tau}\Big)\Big],
\label{eq:W0_model}
\end{equation}. This has shown to reproduce lattice data over a large $\tau$ range in \cite{Bala:2019cqu,Bala:2020tdt}. The idea behind this parametrization is to have a existence of thermal potential, $\lim_{t\to\infty} W(r,\tau\to it)=A e^{-i(V_{re}(r)-i V_{im}(r))t}$. At finite chemical potential in the parametrization above all the coefficients $A_0, V_{re}$ and  $V_{im}$ become chemical potential dependent.
To extract the $\hmu^2$ curvature of the static potential we write,
\begin{align}
V^{Re}(r,\hmu) &= V_0^{Re}(r) + V_2^{Re}(r)/2!\,\hmu^2 + \mathcal{O}(\mu^4), \\
V^{Im}(r,\hmu) &= V_0^{Im}(r) + V_2^{Im}(r)/2!\,\hmu^2 + \mathcal{O}(\mu^4), \\
\ln A(r,\mu) &= \ln A_0(r) + \frac{A_2(r)}{A_0(r)}\,\hmu^2 + \mathcal{O}(\mu^4).
\end{align}
Taking $\partial_\mu^2\ln W|_{\mu=0}$ yields the linear representation
\begin{equation}
Y_2(r,\tau)
= C_0(r)
- V_2^{Re}(r)\,\tau
-\frac{N_\tau V_2^{Im}(r)}{\pi}\ln\sin\!\Big(\frac{\pi\tau}{N_\tau}\Big).
\label{eq:Y2_linear_model}
\end{equation}
Here $C_0(r)=\frac{A_2(r)}{A_0(r)}$. $V_2(r)$ is obtained from a fit to $Y_2(r,\tau)$ as a function of $\tau$.
\section{Flow based subtraction of zero density part of static potential $V_0$}
The potential extracted from the $\hmu=0$ Wilson line correlator contains an additive self energy lattice spacing dependent contribution, which is diverging in the continuum. At finite flow time this is regulated by the flow time and needs to be subtracted for proper renormalization of the potential. In this proceeding, we remove this contribution by subtracting the following term,  
\begin{figure}[t]
    \includegraphics[width=0.44\textwidth]{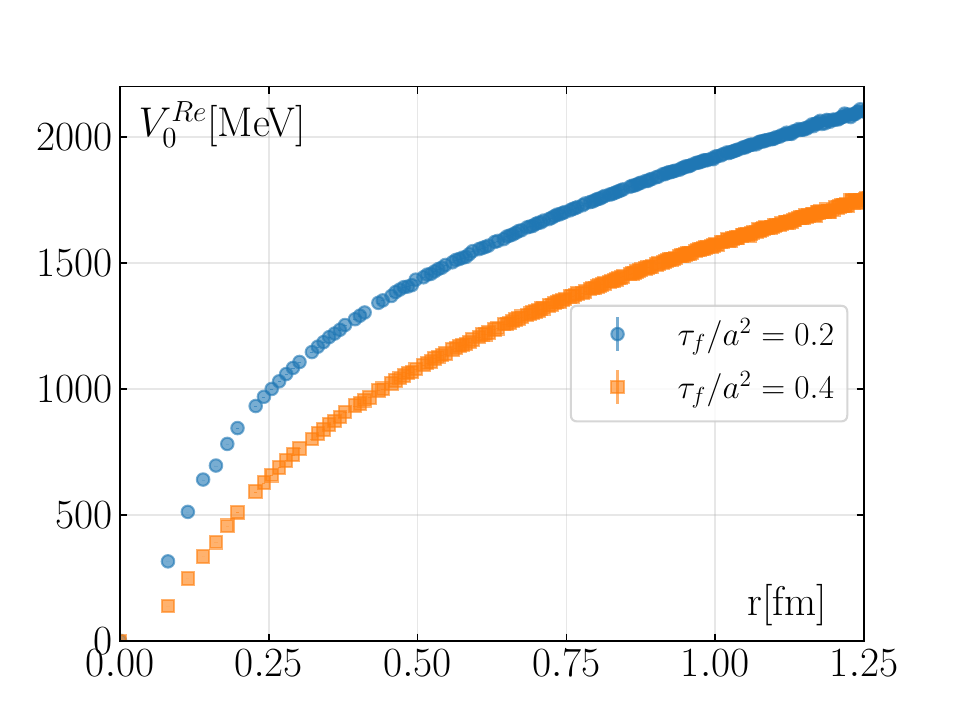}
        \includegraphics[width=0.44\textwidth]{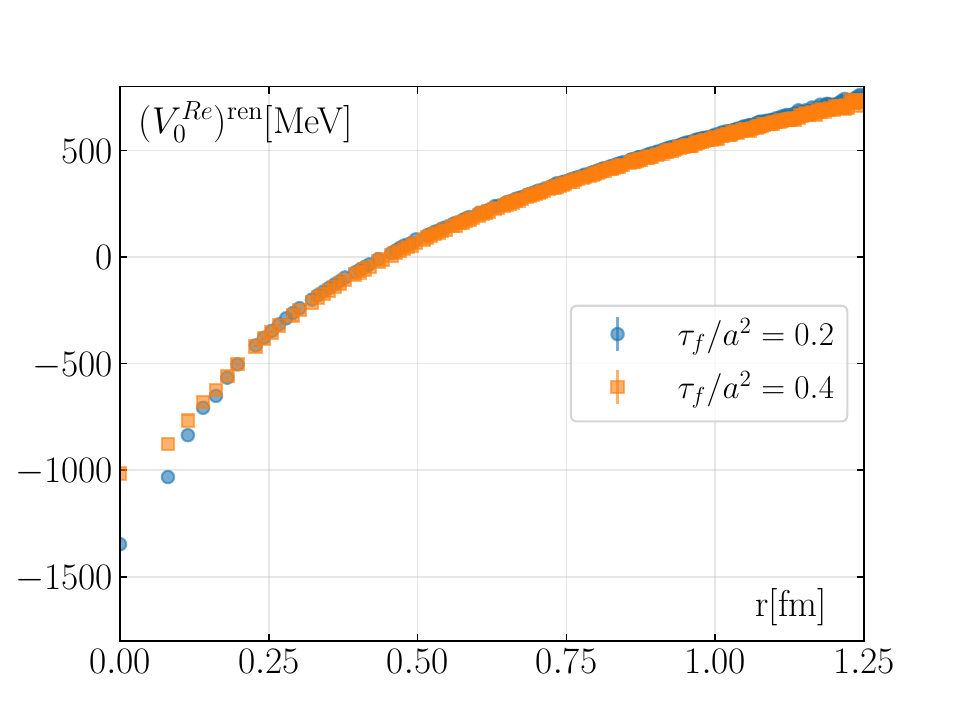}
    \caption{Static thermal potential at vanishing chemical potential for two flow times, $\tau_f/a^2=0.2$, $0.4$, at $T=151.92$ MeV. Left: bare potential $V_0(r)$. Right: renormalized potential
$V_0^{\rm ren}(r)=V_0(r)-\delta V_0(\tau_f/a^2)$ as a function of separation $r$.}
    \label{fig:V_0}
\end{figure}
\begin{equation}
\delta V_0(\tau_f)
=
g^2(\mu_F)\, C_F\, \int_{-\frac{\pi}{a}}^{\frac{\pi}{a}}\frac{d^3k}{(2\pi)^3}\,
\frac{\exp\!\left[-2(\tau_f/a^2)\tilde{k}^2\right]}{\tilde{k}^2},\quad \tilde{k}^2 = 4\sum_{i=1}^3 \sin^2\!\left(\frac{k_i}{2}\right).
\label{eq:deltaV_def}
\end{equation}
This is the zero-temperature diverging self energy contribution in tree-level lattice perturbation theory regulated by flow time. The coupling $g(\mu_F)$ is the $\overline{\text{MS}}$ running coupling evaluated at the scale $\mu_F=\frac{1}{\sqrt{8\tau_f}}$.
The renormalized zero-density potential shown below is then defined by
\begin{equation}
V_0^{\rm ren}(r;\tau_f)=V_0^{\rm bare}(r;\tau_f)-\delta V_0(\tau_f).
\label{eq:V0_ren}
\end{equation}
This improves the consistency between different flow times at intermediate and large separations. At short separations, however, residual finite-flow effects remain visible. A quantitatively controlled determination of the short-distance potential therefore requires an extrapolation to $\tau_f \to 0$.
\begin{figure}[htb]
\centering
\includegraphics[width=0.52\textwidth]{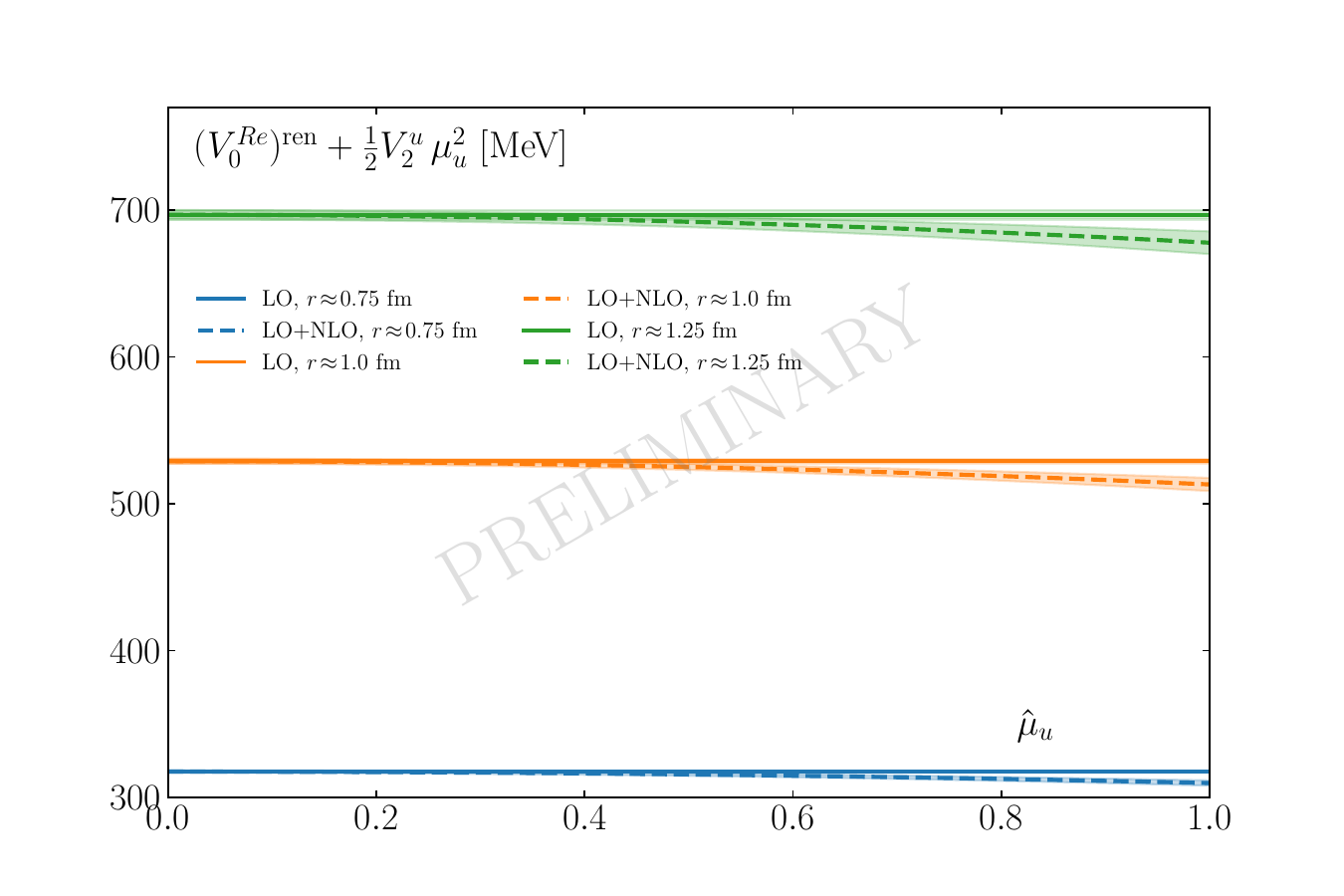}
\caption{Real part of the renormalized static potential in the light quark channel as a function of $\hmu_u$ for selected separations at temperature $T=151.92~\rm{MeV}$ and flow time  $\tau_f/a^2=0.4$. For each separations we compare the leading order(LO) and (LO + NLO) calculations.}
\label{fig:V_mu_total}
\end{figure}

\section{Results}
Fig.~\ref{fig:W2_W0} shows the second order Taylor coefficients of Wilson line correlator, $Y_2^u(r,\tau)$ and $Y_2^s(r,\tau)$, at $T=151.92~\mathrm{MeV}$ for a fixed separation $r=1~\mathrm{fm}$. In both channels, the signal is small at short Euclidean time separation and increases toward larger $\tau$. We also observe a clear channel dependence, with the light-quark contribution larger than the strange-quark contribution. In the following analysis we use the fit interval $\tau\in[3,12)$ and extract both the $\hat{\mu}=0$ and $O(\hat{\mu}^2)$ parts using the ans\"atze given in Eq.(\ref{eq:W0_model}) and (\ref{eq:Y2_linear_model}).
In Fig.(\ref{fig:V_0}) we show the static thermal potential at vanishing density for two flow times $\tau_f/a^2=0.2,0.4$. The left panel shows the bare potential $V_0(r)$, while the right panel shows the renormalized potential. The subtraction removes the flow-time dependent additive constant and at large separations the real potential shows negligible flow time dependence. On the other hand the short distance part has flow-time dependence and a proper zero flow-time extrapolation is necessary. 

The central results of this work are summarized in Fig.~\ref{fig:V_mu_Re} and in Fig.~\ref{fig:V_mu_im}, where we present the $O(\hat{\mu}^2)$ correction to the thermal static potential in the flavor and conserved-charge bases. In both bases, the magnitudes of the real and imaginary parts increase with separation and become clearly visible for $r \gtrsim 0.5~\mathrm{fm}$. This behavior indicates enhanced in-medium screening at finite density. Within our current uncertainties, the $\hat{\mu}^2$ corrections show only a weak dependence on the two flow times considered here. At the largest separations, the statistical errors increase rapidly, but the onset of the finite-density effects remains visible around $r \gtrsim 0.5~\mathrm{fm}$. Moreover, the $\mu^2$ contribution to the imaginary part of the static thermal potential may indicate that finite density not only enhances the in-medium screening but also in-medium broadening relevant for quarkonium dissociation. Fig.~\ref{fig:V_mu_total} illustrates the effect of the $O(\hmu^2)$ correction on the real part of the renormalized static potential in the light-quark channel at flow time $\tau_f/a^2=0.4$. We show selected separations with $r \gtrsim 0.5~\mathrm{fm}$. For all separations considered, the LO+NLO real part of the static potential decreases with increasing chemical potential, and the effect becomes more pronounced toward larger separation $r$. While the correction is small at short distances, it is clearly visible at larger separations and larger $\mu_u/T$.
\section{Summary and outlook}
We presented a lattice QCD study of the static thermal potential at finite density in (2+1)-flavor QCD at $T=151.92~\rm{MeV}$. Using a Taylor expansion of Wilson line correlator around vanishing chemical potentials, we extracted the $\hmu^2$ corrections to the real and imaginary part of the static thermal potential in flavor and conserved charge channels. In all channels, the magnitude of the finite-density correction increases with separation and becomes clearly visible for $r \simeq 0.5,\mathrm{fm}$ indicating enhanced in-medium screening at finite density. While our results are qualitatively consistent with enhanced in-medium screening seen in~\cite{Doring:2005ih,DElia:2019iis} from static quark free energies at non vanishing chemical potentials, our analysis goes beyond these works by extracting the real and imaginary part of the static thermal potential, which are more directly suited for constraining the fate of heavy quarkonium states in the QGP and thus more relevant for heavy ion phenomenology.   Within our current uncertainties, we observe no visible dependence on the two flow times considered in this work. The present analysis is limited by the rapid growth of statistical errors at large separations and by residual flow-time effects at short distances. In the future, we plan to improve this study by increasing the statistics, performing a controlled zero-flow-time extrapolation, estimating the continuum limit, and extending the analysis to more temperatures.

\acknowledgments
This work is supported by the Deutsche Forschungsgemeinschaft, German Research Foundation, Proj. No. 315477589-TRR 211. We sincerely acknowledge the computing time made available on the following : (i) the Noctua2 and Otus clusters at the NHR Center Paderborn, Center for Parallel Computing (PC2), (ii) Bielefeld HPC-GPU cluster at Bielefeld University. We also thank Frithjof Karsch for useful discussions. We thank the HotQCD collaboration for sharing the gauge configurations. We use the SIMULATeQCD package for generating the gauge configurations and also measuring the Wilson line correlators ~\cite{HotQCD:2023ghu}. For dataanalysis we have used~\cite{toolbox}.
\bibliographystyle{JHEP}
\bibliography{bibliography}

\providecommand{\href}[2]{#2}\begingroup\raggedright\begin{thebibliography}{10}

\bibitem{Matsui:1986dk}
T.~Matsui and H.~Satz, \emph{{$J/\psi$ Suppression by Quark-Gluon Plasma
  Formation}}, \href{https://doi.org/10.1016/0370-2693(86)91404-8}{\emph{Phys.
  Lett. B} {\bfseries 178} (1986) 416}.

\bibitem{Laine:2006ns}
M.~Laine, O.~Philipsen, P.~Romatschke and M.~Tassler, \emph{{Real-time static
  potential in hot QCD}},
  \href{https://doi.org/10.1088/1126-6708/2007/03/054}{\emph{JHEP} {\bfseries
  03} (2007) 054} [\href{https://arxiv.org/abs/hep-ph/0611300}{{\ttfamily
  hep-ph/0611300}}].

\bibitem{Burnier:2007qm}
Y.~Burnier, M.~Laine and M.~Vepsalainen, \emph{{Heavy quarkonium in any channel
  in resummed hot QCD}},
  \href{https://doi.org/10.1088/1126-6708/2008/01/043}{\emph{JHEP} {\bfseries
  01} (2008) 043} [\href{https://arxiv.org/abs/0711.1743}{{\ttfamily
  0711.1743}}].

\bibitem{Brambilla:2008cx}
N.~Brambilla, J.~Ghiglieri, A.~Vairo and P.~Petreczky, \emph{{Static
  quark-antiquark pairs at finite temperature}},
  \href{https://doi.org/10.1103/PhysRevD.78.014017}{\emph{Phys. Rev. D}
  {\bfseries 78} (2008) 014017}
  [\href{https://arxiv.org/abs/0804.0993}{{\ttfamily 0804.0993}}].

\bibitem{Burnier:2014ssa}
Y.~Burnier, O.~Kaczmarek and A.~Rothkopf, \emph{{Static quark-antiquark
  potential in the quark-gluon plasma from lattice QCD}},
  \href{https://doi.org/10.1103/PhysRevLett.114.082001}{\emph{Phys. Rev. Lett.}
  {\bfseries 114} (2015) 082001}
  [\href{https://arxiv.org/abs/1410.2546}{{\ttfamily 1410.2546}}].

\bibitem{Burnier:2015tda}
Y.~Burnier, O.~Kaczmarek and A.~Rothkopf, \emph{{Quarkonium at finite
  temperature: Towards realistic phenomenology from first principles}},
  \href{https://doi.org/10.1007/JHEP12(2015)101}{\emph{JHEP} {\bfseries 12}
  (2015) 101} [\href{https://arxiv.org/abs/1509.07366}{{\ttfamily
  1509.07366}}].

\bibitem{Bala:2019cqu}
D.~Bala and S.~Datta, \emph{{Nonperturbative potential for the study of
  quarkonia in QGP}},
  \href{https://doi.org/10.1103/PhysRevD.101.034507}{\emph{Phys. Rev. D}
  {\bfseries 101} (2020) 034507}
  [\href{https://arxiv.org/abs/1909.10548}{{\ttfamily 1909.10548}}].

\bibitem{Ali:2025iux}
{\scshape HotQCD} collaboration, \emph{{Thermal static potential and
  pseudoscalar quarkonium spectral functions from (2+1)-flavor lattice QCD}},
  \href{https://doi.org/10.1103/s8gw-n43f}{\emph{Phys. Rev. D} {\bfseries 112}
  (2025) 054510} [\href{https://arxiv.org/abs/2505.11313}{{\ttfamily
  2505.11313}}].

\bibitem{Bala:2026vnl}
D.~Bala, \emph{{Some results on heavy-flavor systems at finite temperature from
  lattice QCD}}, \href{https://doi.org/10.1016/j.jspc.2026.100288}{\emph{J.
  Subatomic Part. Cosmol.} {\bfseries 5} (2026) 100288}.

\bibitem{Gavai:2008zr}
R.~V. Gavai and S.~Gupta, \emph{{QCD at finite chemical potential with six time
  slices}}, \href{https://doi.org/10.1103/PhysRevD.78.114503}{\emph{Phys. Rev.
  D} {\bfseries 78} (2008) 114503}
  [\href{https://arxiv.org/abs/0806.2233}{{\ttfamily 0806.2233}}].

\bibitem{Bazavov:2017dus}
A.~Bazavov et~al., \emph{{The QCD Equation of State to $\mathcal{O}(\mu_B^6)$
  from Lattice QCD}},
  \href{https://doi.org/10.1103/PhysRevD.95.054504}{\emph{Phys. Rev. D}
  {\bfseries 95} (2017) 054504}
  [\href{https://arxiv.org/abs/1701.04325}{{\ttfamily 1701.04325}}].

\bibitem{Bollweg:2022rps}
{\scshape HotQCD} collaboration, \emph{{Taylor expansions and Pad{\'e}
  approximants for cumulants of conserved charge fluctuations at nonvanishing
  chemical potentials}},
  \href{https://doi.org/10.1103/PhysRevD.105.074511}{\emph{Phys. Rev. D}
  {\bfseries 105} (2022) 074511}
  [\href{https://arxiv.org/abs/2202.09184}{{\ttfamily 2202.09184}}].

\bibitem{Bollweg:2022fqq}
{\scshape HotQCD} collaboration, \emph{{Equation of state and speed of sound of
  (2+1)-flavor QCD in strangeness-neutral matter at nonvanishing net
  baryon-number density}},
  \href{https://doi.org/10.1103/PhysRevD.108.014510}{\emph{Phys. Rev. D}
  {\bfseries 108} (2023) 014510}
  [\href{https://arxiv.org/abs/2212.09043}{{\ttfamily 2212.09043}}].

\bibitem{HotQCD:2018pds}
{\scshape HotQCD} collaboration, \emph{{Chiral crossover in QCD at zero and
  non-zero chemical potentials}},
  \href{https://doi.org/10.1016/j.physletb.2019.05.013}{\emph{Phys. Lett. B}
  {\bfseries 795} (2019) 15}
  [\href{https://arxiv.org/abs/1812.08235}{{\ttfamily 1812.08235}}].

\bibitem{Doring:2005ih}
M.~Doring, S.~Ejiri, O.~Kaczmarek, F.~Karsch and E.~Laermann, \emph{{Screening
  of heavy quark free energies at finite temperature and non-zero baryon
  chemical potential}},
  \href{https://doi.org/10.1140/epjc/s2005-02462-y}{\emph{Eur. Phys. J. C}
  {\bfseries 46} (2006) 179}
  [\href{https://arxiv.org/abs/hep-lat/0509001}{{\ttfamily hep-lat/0509001}}].

\bibitem{Allton:2005gk}
C.~R. Allton, M.~Doring, S.~Ejiri, S.~J. Hands, O.~Kaczmarek, F.~Karsch et~al.,
  \emph{{Thermodynamics of two flavor QCD to sixth order in quark chemical
  potential}}, \href{https://doi.org/10.1103/PhysRevD.71.054508}{\emph{Phys.
  Rev. D} {\bfseries 71} (2005) 054508}
  [\href{https://arxiv.org/abs/hep-lat/0501030}{{\ttfamily hep-lat/0501030}}].

\bibitem{Bala:2020tdt}
D.~Bala and S.~Datta, \emph{{Interaction potential between heavy
  $Q\overline{Q}$ in a color octet configuration in the QGP from a study of
  hybrid Wilson loops}},
  \href{https://doi.org/10.1103/PhysRevD.103.014512}{\emph{Phys. Rev. D}
  {\bfseries 103} (2021) 014512}
  [\href{https://arxiv.org/abs/2009.00773}{{\ttfamily 2009.00773}}].

\bibitem{DElia:2019iis}
M.~D'Elia, F.~Negro, A.~Rucci and F.~Sanfilippo, \emph{{Dependence of the
  static quark free energy on $\mu_B$ and the crossover temperature of $N_f =
  2+1$ QCD}}, \href{https://doi.org/10.1103/PhysRevD.100.054504}{\emph{Phys.
  Rev. D} {\bfseries 100} (2019) 054504}
  [\href{https://arxiv.org/abs/1907.09461}{{\ttfamily 1907.09461}}].

\bibitem{HotQCD:2023ghu}
{\scshape HotQCD} collaboration, \emph{{SIMULATeQCD: A simple multi-GPU lattice
  code for QCD calculations}},
  \href{https://doi.org/10.1016/j.cpc.2024.109164}{\emph{Comput. Phys. Commun.}
  {\bfseries 300} (2024) 109164}
  [\href{https://arxiv.org/abs/2306.01098}{{\ttfamily 2306.01098}}].

\bibitem{toolbox}
L.~Altenkort, D.~A. Clarke, J.~Goswami and H.~Sandmeyer, \emph{{Streamlined
  data analysis in Python}}, {\emph{PoS} {\bfseries LATTICE2023} (2023) 136}
  [\href{https://arxiv.org/abs/2308.06652}{{\ttfamily 2308.06652}}].

\end{thebibliography}\endgroup
\end{document}